\documentstyle[amssymb,twocolumn,prl,aps]{revtex}
\makeatletter
\renewcommand\@cite[1]{(#1)}
\makeatother
\makeatletter
\renewcommand\@biblabel[1]{#1.}
\makeatother
\begin{document}

\onecolumn
\def\lscu{La$_{2-x}$Sr$_{x}$CuO$_{4}$\ }
\def\lsco1{La$_{1.837}$Sr$_{0.163}$CuO$_{4}$\ }
\def\ybco{YBa$_{2}$Cu$_{3}$O$_{7-y}$\ }
\newcommand{\lapprox}{\stackrel{<}{\scriptstyle \sim}}
\newcommand{\gapprox}{\stackrel{>}{\scriptstyle \sim}}
\title{
\vspace{2cm}
\large
\bf Spins in the Vortices of a High Temperature Superconductor \\
\vspace{0.8cm} }

\author{B.\ Lake$^{1}$, G.\ Aeppli$^{2,3}$, K.\ N.\ Clausen$^{3}$, 
D.\ F.\ McMorrow$^{3}$, K.\ Lefmann$^{3}$, 
N.\ E.\ Hussey$^{4,5}\footnote{Present address H.H. Wills Physics Laboratory, 
University of Bristol, Tyndall Avenue, Bristol, BS8 1TL, U.K.}$, 
N.\ Mangkorntong$^{4}$ M.\ Nohara$^{4}$, H.\ Takagi$^{4}$, 
T.\ E.\ Mason$^{1}$, and A.\ Schr\"{o}der$^{6}$.}

\address{
$^{1}$Oak Ridge National Laboratory, Oak Ridge, Tennessee 37831, U.S.A.  \\
$^{2}$N.E.C Research, 4 Independence Way, Princeton, New Jersey 08540, 
U.S.A. \\
$^{3}$Department of Condensed Matter Physics and Chemistry, 
Ris\o\ National Laboratory, 4000 Roskilde, Denmark \\
$^{4}$Department of Advanced Material Science, 
Graduate School of Frontier Sciences, University of Tokyo, 
Hongo7-3-1, Bunkyo-ku, Tokyo 113-8656, Japan.  \\
$^{5}$Department of Physics, University of Loughborough, Loughborough,
LE11 3TU, U.K. \\
$^{6}$Physikalisches Institute, University of Karlsruhe, D-76128 Karlsruhe,
Germany \\ }


\maketitle


\begin{abstract}

Neutron scattering is used to characterise the magnetism of the 
vortices for the optimally doped high-temperature superconductor
\lscu ($x=0.163$) in an applied magnetic field. As temperature is 
reduced, low frequency spin fluctuations first disappear with the 
loss of vortex mobility, but then reappear. We find that the vortex 
state can be regarded as an inhomogeneous mixture of a superconducting 
spin fluid and a material containing a nearly ordered antiferromagnet.
These experiments show that as for many other properties of cuprate 
superconductors, the important underlying microscopic forces are 
magnetic.
\end{abstract}

\pacs{PACS numbers: 74.72.Dn, 61.12.Ex Hk, 74.25.Ha}

\newpage

Many of the practical applications of type-II superconductors, rely 
on their ability to carry electrical currents without dissipation even 
in magnetic fields greater than the Meissner field, below which the 
superconductor excludes magnetic flux entirely. At such high magnetic 
fields, superconductors are in a mixed state or 'vortex lattice', 
comprising an array of cylindrical inclusions (vortices) of normal 
material in a superconducting matrix. Vortex lattices have two magnetic 
aspects. The first is that there are magnetic field gradients due to the 
inhomogeneous flux penetration - each vortex allows a magnetic flux 
quantum to penetrate and the magnetic field decays from the vortex center 
into the superconductor over a distance of order the London length 
$\lambda$ (\emph{1}). $\lambda$ is the depth beyond which the 
superconductor excludes small fields, and is typically between 100 and 
1000 nm. The second is that the electron spins in the non-superconducting 
cores should no longer be paired coherently (as they are in the 
superconducting state). The length scale for this microscopic magnetic 
effect is the radius $\xi$ of the Cooper pairs which 
underlie the phenomenon of superconductivity. $\xi$ generally ranges 
from 100 nm - common for conventional, low transition temperature 
($T_{c}$) superconductors - down to nearly a nm, which is found for the 
high-$T_{c}$ copper oxides. Most magnetic measurements of the flux 
lattice state, including images from neutron diffraction and microscopy 
of magnetic nanoparticles deposited on samples threaded by vortices, 
are sensitive primarily to the mesoscopic field gradients characterized 
by $\lambda$ (\emph{2-5}). Much less is known 
about the microscopic magnetism of the vortices. The associated spin 
correlations and dynamics are important because they mirror the internal 
structure of the vortices and in superconductors with strong magnetic 
interactions, are likely to dominate vortex state energetics and 
thermodynamics. Thus motivated, we have performed an experiment which 
images the spin correlations in the vortex state of the simplest 
high-temperature superconductor, \lscu.
The key finding is 
that the vortex state for our optimally doped sample ($x=0.163$, 
superconducting transition temperature $T_{c}=38.5$ K) has much stronger 
tendencies towards magnetic order than either the normal or superconducting 
states. 

We used inelastic neutron scattering to measure $\chi''$ (the Fourier 
transform of the two-spin correlations divided by the Bose factor) as a 
function of momentum and energy. The superconducting CuO$_{2}$ planes of our 
sample were placed in the horizontal scattering plane of a neutron scattering 
spectrometer, and the magnetic field $H$ was applied perpendicular to these 
planes in the vertical direction\footnote{The single crystal samples were 
described previously (\emph{6}), and 11 of them (with a total weight 
of 25 grams) were mutually aligned to within $\pm 0.6$ and $\pm 3.9$ 
degrees in directions parallel and perpendicular to the CuO$_{2}$ basal 
planes respectively. The resulting sample was placed on the cold finger of 
a variable temperature insert in a split coil superconducting magnet, which 
in turn was installed on the sample table of the RITA (TAS6) cold neutron 
spectrometer at DR3, Ris\o\ National Laboratory (\emph{7,8}).}. A 
sliver of one crystal was used for magnetotransport measurements, and these 
were combined with earlier data for $x=0.17$ (\emph{9}) to establish the 
$H-T$ phase diagram (Fig.\ 1A). The electrical resistance vanishes below an 
irreversibility line (\emph{10}) which is a very rapid function of 
applied field, so that even for fields well below the upper critical 
field $H_{c2}$ (defined here as the field at which non-zero resistivity 
is first detected), the vortex lattice required for macroscopic 
superconducting phase coherence and perfect conductivity does not occur 
until $T$ is well below $T_{c}(H=0)=38.5$ K, the zero field transition 
temperature.

\lsco1 is characterised by spin fluctuations occuring at a quartet of 
$x$-dependent characteristic wavevectors given by
$\mbox{\boldmath $Q$}_{\delta}=(\frac{1}{2}(1\pm\delta),\frac{1}{2})$ and 
$(\frac{1}{2},\frac{1}{2}(1\pm\delta))$ with $\delta=0.254$ 
(\emph{11,12}) (Fig. 1B). 
Superconductivity has several effects on the magnetic fluctuations, the 
most pronounced of which is that an energy gap, $\Delta$, appears in the 
spectrum (\emph{6,13}) (Fig.\ 2A). On cooling from $T_{c}$=38.5 K 
to 5.5 K in zero field, the normal state continuum is eliminated below 
$\Delta$=6.7 meV. Application of a 7.5 T field fills the gap 
at base temperature with a spectrum 
whose amplitude is little different from that seen in the same energy 
range in zero field at $T_{c}$. This result means that the vortex state 
for a field far below the upper critical field $H_{c2}\approx62$ T 
(\emph{9}), where the vortex cores presumably occupy a small volume fraction 
of the material ($H/H_{c2}$=12\%), displays low-frequency magnetic 
fluctuations of roughly the same strength as the ungapped normal state. 
The difference plot (Fig.\ 2B) shows that the field-induced signal peaks 
at $4.3 \pm 0.5$ meV, and dwindles to zero as $E$ approaches either 0 or 
$\Delta$. 
This characteristic energy is approximately half that of the normal state 
indicating that the fluctuations in the field are two times slower 
and have a greater tendancy towards antiferromagnetic order.

To understand the spatial nature of the field-induced sub-gap fluctuations, 
we have performed scans for fixed energy transfer as a function of 
wavevector along the solid black trajectory shown Fig.\ 1B. At low 
temperatures and energies below the gap ($T=6.6$ K, $E$=2.5 meV) sharp peaks 
are observed in a field of $H$=7.5 T (Fig. 3A) at the same wavevectors where 
the normal state response is maximal (Fig.\ 3B). We also investigated the 
magnetic correlations just above the spin-gap at 7.5 meV and find that 
the field has no discernable effect on the magnetic correlations at this 
energy (Fig.\ 3C and D).

Closer examination of Figs\ 3A-D yields a wealth of quantitative information 
about the microscopic magnetism of the vortex state. First, the preferred 
periodicity, derived from the peak positions $\mbox{\boldmath $Q$}_{\delta}$
of the magnetization density, is $1/\delta = (3.93\pm0.09)a_{0}$, where 
$a_{0}=3.777$ $\AA$ is the Cu$^{2+}$-Cu$^{2+}$ separation. This value is 
indistinguishable from the periodicity found for the normal state. Second,
the scattering profile is slightly different from that measured at 38.5 K 
for $H=0$, in that the peaks seem sharper in the vortex state even though 
the scattering between the peaks has the same amplitude when scaled to the 
peaks. Indeed the peaks are as sharp as the instrumental resolution permits,
implying that the principal (period $1/\delta$) magnetization oscillations 
in the vortex state are coherent over distances $l_{v}>20a_{0}$, to be 
compared to distances $l_{n}=(6.32\pm0.22)a_{0}$ for the normal state 
(\emph{6}). For comparison, the lattice constant for a well-formed 
(Abrikosov) vortex state is $a_{V}=(2\Phi_{0}/\sqrt{3}H)^{\frac{1}{2}}$ 
where the magnetic flux quantum $\Phi_{0}=2067$ T(nm)$^{2}$. At 7.5 T, 
$a_{V}= 47.2a_{o}$, a number much larger than $1/\delta$, but potentially 
similar to $l_{v}$.

We have also measured the temperature dependence of the field-induced 
response. Fig.\ 4 shows electrical resistivity as well as neutron data, 
collected with wavevector and energy fixed at $\mbox{\boldmath $Q$}_{\delta}$ 
and 2.5 meV respectively. At $H=0$, the neutron signal undergoes a sharp drop 
starting at $T_{c}$ (Fig.\ 4C), (which is where the transition to zero 
resistance also occurs (Fig.\ 4A)) and it dwindles into the background 
below 15 K. A field of $H=7.5$ T has a large effect on the temperature 
evolutions of both the resistivity and neutron intensity. The resistivity 
descends steadily as temperature decreases between $T_{c}$ and 30 K 
(Fig.\ 4A), and does not have its final inflection point, as measured by 
$d\rho_{ab}/dT$ (Fig.\ 4B) until 25K, which is also where irreversibility 
sets in. This inflection point has been found (\emph{14}) to coincide 
with the drop in the magnetization associated with the freezing transition 
of vortices - above the freezing point, the imposed current loses energy 
via vortex motion, while below, the vortices are pinned and the current 
is dissipationless. The corresponding magnetic neutron scattering signal 
(Fig 4C), which is slightly suppressed at $T>T_{c}$, remains close to its 
normal state value for $T_{c}>T>25$ K, and undergoes a sharp decline below 
25 K. Thus, our spin signal, a microscopic probe of vortices, tracks a 
macroscopic measure - namely the electrical resistance - of vortex freezing, 
which in layered materials such as the cuprates can occur well below 
$T_{c}$ even for $H \ll H_{c2}$, the upper critical field (\emph{15}). 

How can our microscopic results be connected to the bulk data? In the 
vortex fluid state for $T>25$ K, all Cu$^{2+}$ sites are visited 
occasionally by vortices and then depart. While at the sites, they 
establish a decaying (in time) magnetization density wave, the quantity 
which our experiment is sensitive to. Thus all Cu$^{2+}$ sites would 
have some memory of visits by vortices. When the inverse residence time 
$\tau^{-1}$ of a vortex at any site approaches the frequency of the 
measured spin fluctuations the fraction of sites with such memory 
will begin to significantly exceed the fraction $H/H_{c2}$ of sites 
covered by vortices at a given instant. The resistivity data in Fig. 4A 
yield the crude estimate 2.5 meV for $\hbar\tau^{-1}$ at 30 K (\emph{16}), 
which happens to coincide with $\hbar\omega$ in Fig. 4C. As $T$ is 
cooled below 25 K, the vortices become pinned via a combination of their 
mutual interactions and intrinsic disorder, 
so that they are always present at certain sites and never present at 
others. The outcome is then that subgap magnetization fluctuations occur 
only near the relatively small fraction of sites where vortices are pinned, 
with the result that the magnetic response is correspondingly reduced.

Although observing vortex freezing via the electron spin correlations is 
unprecedented, an even more fascinating phenomenon occurs below 10 K. Here, 
the decline of the signal below the freezing transition is reversed, 
resulting in a susceptibility approximately equal to the normal state 
$\chi''$. Macroscopic measurements (\emph{9}) do not indicate any changes 
in the vortex order or dynamics. Therefore, while they can plausibly account 
for the abruptly falling signal near 25 K, such changes cannot be responsible 
for the rising signal below 10 K. We conclude that the low-$T$ increase 
can only follow from changes in the magnetism of the frozen vortex 
matter, and speculate that its most likely cause cannot be the relatively 
large magnetic interactions we suspect exsist within individual vortices, 
but rather the weaker interactions between spins in different vortices that 
become relevant only at low-$T$.

Our data show that a modest field induces extraordinary subgap
excitations in the optimally doped high-$T_{c}$ superconductor \lsco1.
There are several possible origins for such excitations. The first are
the quasiparticles inhabiting the vortex cores, which in
conventional superconductors are simply metallic tubes with finite-size
quantization of electron orbits perpendicular to the tube axes
(\emph{17,18}). The second, felt to be responsible for the
$\sqrt{H}$ low-$T$ specific heat in $d$-wave superconductors, is due to the
nodal quasiparticles whose energies are Doppler-shifted by the
supercurrents around the vortices (\emph{19}). The third is that the
cores are small antiferromagnets, but that because of finite size
quantization, and the weak magnetic interactions between planes as well
as between vortices within the same planes, the antiferromagnetic
correlations are dynamic and so are characterized by finite oscillation 
frequencies and relaxation rates. The first are excluded because
the signal which we measure is comparable to that found in
the normal state, giving a superconducting to normal state signal ratio
which is much larger than the volume fraction $H/H_{c2}$ occupied by vortices
in such models. The second has an analogous difficulty with the
low-$T$ specific heat $C$, which does appear to follow the $d$-wave
prescription for our samples (\emph{20}), in that the ratio of the
low temperature (for 7.5 T) and paramagnetic phase Sommerfeld constants
$C/T$ is 15 \%, and is therefore much less than the ~100 \% ratio of 
field-induced low-$T$ to
zero-field paramagnetic signals measured with the neutrons. This leaves
us with the third option, where we imagine the vortex state as an
inhomogeneous mixture not only of paramagnetic and superconducting regions, 
but as a magnetically inhomogeneous mixture as well. The superconducting 
regions have a well-defined spin gap, while the paramagnetic regions 
contain fluctuations towards long period magnetic order. This picture 
accounts better for the observed subgap spectral weight than the other 
two scenarios. Specifically, we estimate that the net subgap weight, (placed 
in absolute units using normalization to phonon scattering (\emph{21}))
integrated over energy and reciprocal space, corresponds to
$0.05\mu_{B}^{2}$/Cu$^{2+}$, which is remarkably close to 
$H/H_{c2}\mu_{2D,S=\frac{1}{2}}^{2}=0.044\mu_{B}^{2}$, the product of 
the volume fraction occupied by the vortices and the square of the ordered 
moment $\mu_{2D,S=\frac{1}{2}}=0.6\mu_{B}$ found in insulating 
two-dimensional $S$=1/2 Heisenberg antiferromagnets (\emph{22}). In other 
words, the ordered moment which for the model insulator appears as an 
elastic Bragg peak, becomes a fluctuating moment manifested in the 
inelastic subgap peak for the vortex state of the superconductor.

While the simple picture of inclusions of finite-size vortices with
large spin density wave susceptibilities accounts for many of our
observations, the material in a field cannot be simply visualized as a
superconductor perforated by an array of independent, nearly
antiferromagnetic cylinders with diameter given by the pair coherence
length. First, the spin density period is of order the pair coherence
length $\xi$, and the magnetic correlation length is substantially longer
than $\xi$. Second, the magnetic field also induces broad scattering 
between the incommensurate peaks, with a characteristic length scale 
of order $a_{0}\ll\xi$. Third, as described above, the low-$T$ rise in 
$\chi''$ is difficult to explain without invoking weak interactions 
between the spins in 'separate' vortices. The observations together show 
that the spins in the vortices are correlated over a variety of length 
scales from the atomic to the mesoscopic. The most natural explanation 
is that the vortices themselves are highly anisotropic objects, or at 
the very least, have a highly anisotropic effect on the spin correlations 
in the intervening superconducting region. Such anisotropy is quite 
consistent with a $d$-wave pairing state, although it has not been found 
in scanning probe images of vortices of high-$T_{c}$ superconductors 
(\emph{23,24}).

We have measured the microscopic spin correlations associated with the 
vortex state in the optimally doped single-layer high-$T_{c}$ cuprate 
La$_{1.84}$Sr$_{0.16}$CuO$_{4}$. We discover that on cooling in a modest 
field, low energy spin fluctuations are suppressed not near the zero-field
transition, but at the irreversibility line below which the
superconductor is in a true zero resistance state. This links the
development of the spin gap more to superconducting phase coherence - 
required for zero electrical resistance - throughout the sample than to 
local pairing. A second discovery is that at low temperatures, the vortex 
matter exhibits a rising tendency towards the magnetic order found for the 
'striped' state (\emph{25-27}) with $x$=1/8. This notion 
is in broad agreement with theoretical ideas (\emph{28-30}) and 
implies that in the $H$-$T$ plane, the critical line separating frozen 
from fluid-like 'vortex' states may actually mark mesoscopic phase 
separation - or 'gellation' - into a nearly magnetic insulating vortex 
network bathed in a superconducting quantum spin fluid.


\newpage

\begin{figure}
\caption{
Phase diagram and wavevector map of \lsco1. (A) Shows the red 
irreversibility line in the $H-T$ plane which separates the resistive 
normal/vortex fluid state from the superconducting state. Red circles 
come from the magneto-transport measurements (shown in Fig.\ 4A) and 
mark the temperatures where non-zero resistivity is first detected for 
a given field. We also show the data (red squares) of Ando and coworkers 
(\emph{9}) for an $x$=0.17 sample. The blue arrow represents the trajectory 
of the $H=7.5$ T temperature scan shown in Fig.\ 4C. (B) shows reciprocal 
space for the superconducting CuO$_{2}$ planes of \lsco1, as probed by our 
neutron scattering measurements. Spin fluctuations are observed at a 
quartet of incommensurate wavevectors indicated by the red dots. The solid 
black arc shows the wavevectors measured in a typical constant-energy scan 
(see Fig.\ 3) and the green ellipse represents the instrumental resolution. 
The magnetic field was applied perpendicular to the CuO$_{2}$ planes, and 
the blue dots indicate the reciprocal lattice associated with the $H=7.5$ T 
vortex state. }
\end{figure}

\begin{figure}
\caption{
Constant-wavevector scans plotted as functions of energy. The  energy 
resolution is 0.41 meV full-width-at-half-maximum. (A) shows the magnetic 
susceptibility $\chi ''$, measured at the incommensurate peak 
$\mbox{\boldmath $Q$}_{\delta}$, in zero applied field for both the 
superconducting state (red circles) and the normal state (red triangles). 
$\chi ''$ was also measured in a magnetic field of $H=7.5$ T at $T=7.7$ K 
(blue circles). The blue line through the data in {\bf a} corresponds to 
the damped harmonic oscillator 
$E_{o}^{2}E\gamma/((E^{2}-E_{o}^{2})^{2}+E^{2}\gamma^{2})$ (with 
$E_{o}=4.3 \pm 0.5$ meV and $\gamma=4.3 \pm 0.2$ meV), which models the 
magetism of the vortices plus the gapped form described in Ref.\ (\emph{6}), 
to account for the remaining superconducting signal. The dashed red 
line is the form from Ref.\ (\emph{6}) alone and describes the gapped 
spin-fluid-like response of the superconductor at $H=0$ T. The dotted 
red line is the quasi-elastic response $E\Gamma/(E^{2}+\Gamma^{2})$ (with 
$\Gamma = 9$ meV), that was used previously to account for the normal 
state signal (\emph{6}). 
(B) shows the 
difference between the $H=7.5$ T and $H=0$ results at low $T$. The 
solid blue line is simply the difference between the blue and dashed red 
lines in (A). } 
\end{figure}

\begin{figure}
\caption{
Constant-energy scans plotted as a function of wavevector along the black 
trajectory shown in Fig.\ 1B. (A) shows the susceptibility measured for 
$T=6.6$ K, below the energy gap. The data are sums of scans for $E=1.5$, 
2.5 and 3.5 meV. In zero field (red circles) the susceptibility is 
completely suppressed by superconductivity and application of a 7.5 T 
field (blue circles) induces a subgap signal. For comparison the normal 
state susceptibility is shown in (B). (C) and (D) give the susceptibility 
above the energy gap at $E=7.5$ meV, in both the normal and superconducting 
states respectively. 
The lines in all frames except (A) are the resolution-corrected Sato-Maki 
lineshape (\emph{31}); for $H=0$, the width parameters were fixed at the 
values established from the higher resolution data of Ref. (\emph{6}).
The solid blue line in (A) is the fit to the normal state data from (B), 
scaled to match the peak amplitudes, while the dashed blue line consists 
of two peaks representing the resolution of the instrument. 
}
\end{figure}

\begin{figure}
\caption{ Temperature-dependent electrical transport and neutron data. 
(A) Shows the in-plane resistivity data collected at a variety of fields 
from $H=0$ to 9 T. 
(B) gives the derivative of the in-plane resistivity with respect to 
temperature at $H=7.5$ T, which is the field employed in the neutron 
scattering experiment. (C) shows the magnetic susceptibilities $\chi"$ at 
$\mbox{\boldmath $Q$}=\mbox{\boldmath $Q$}_{\delta}$ and below the energy 
gap at $E=2.5$ meV, for $H=0$ (red circles) and 7.5 T (blue circles). 
}
\end{figure}

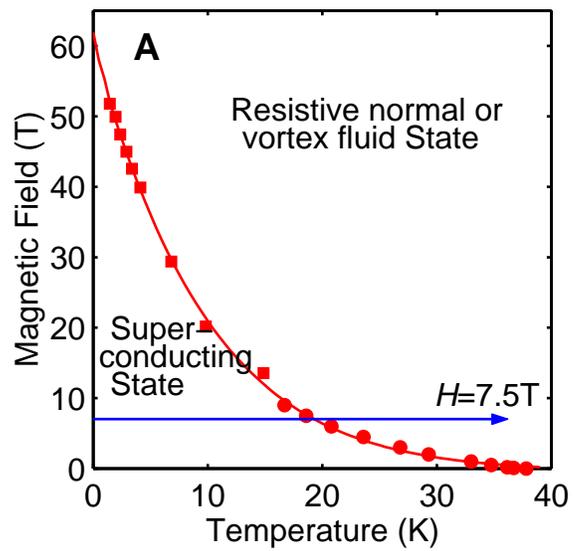
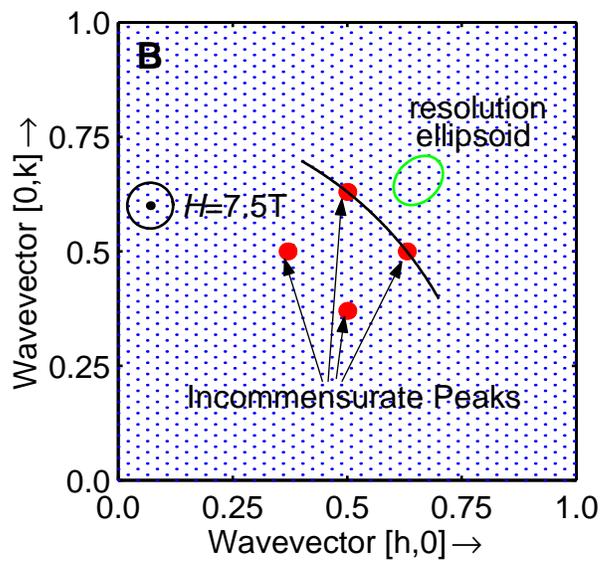
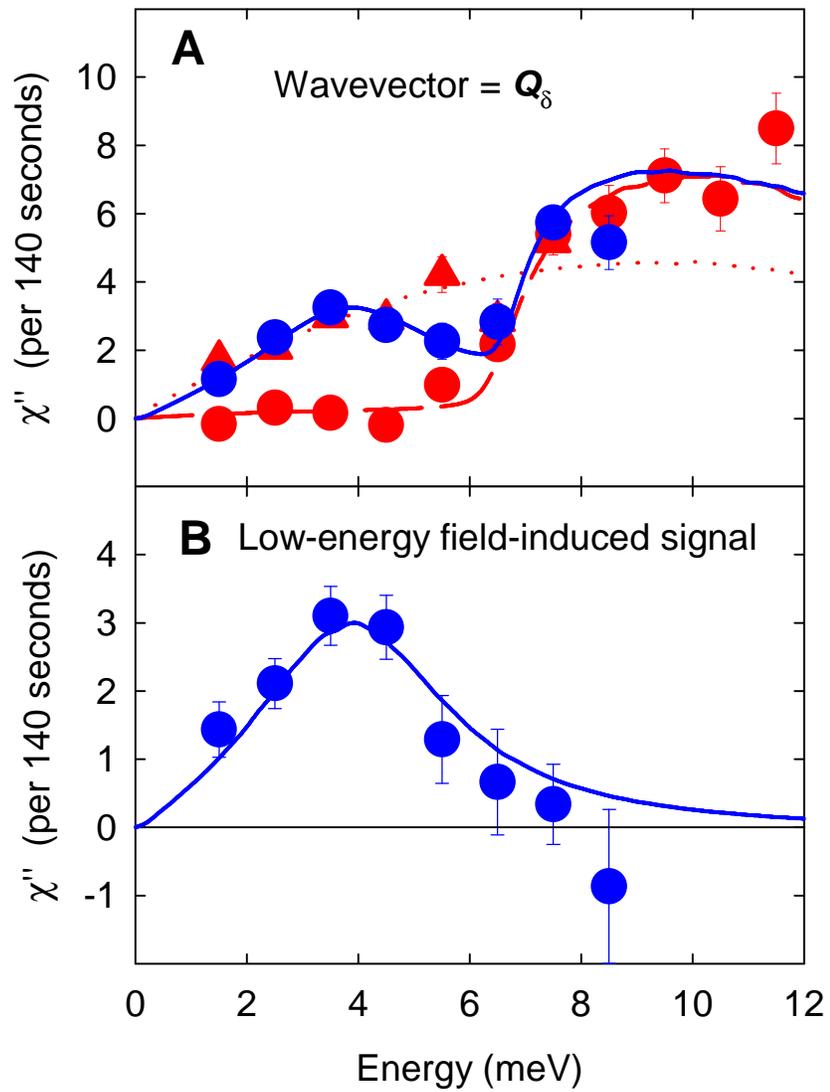
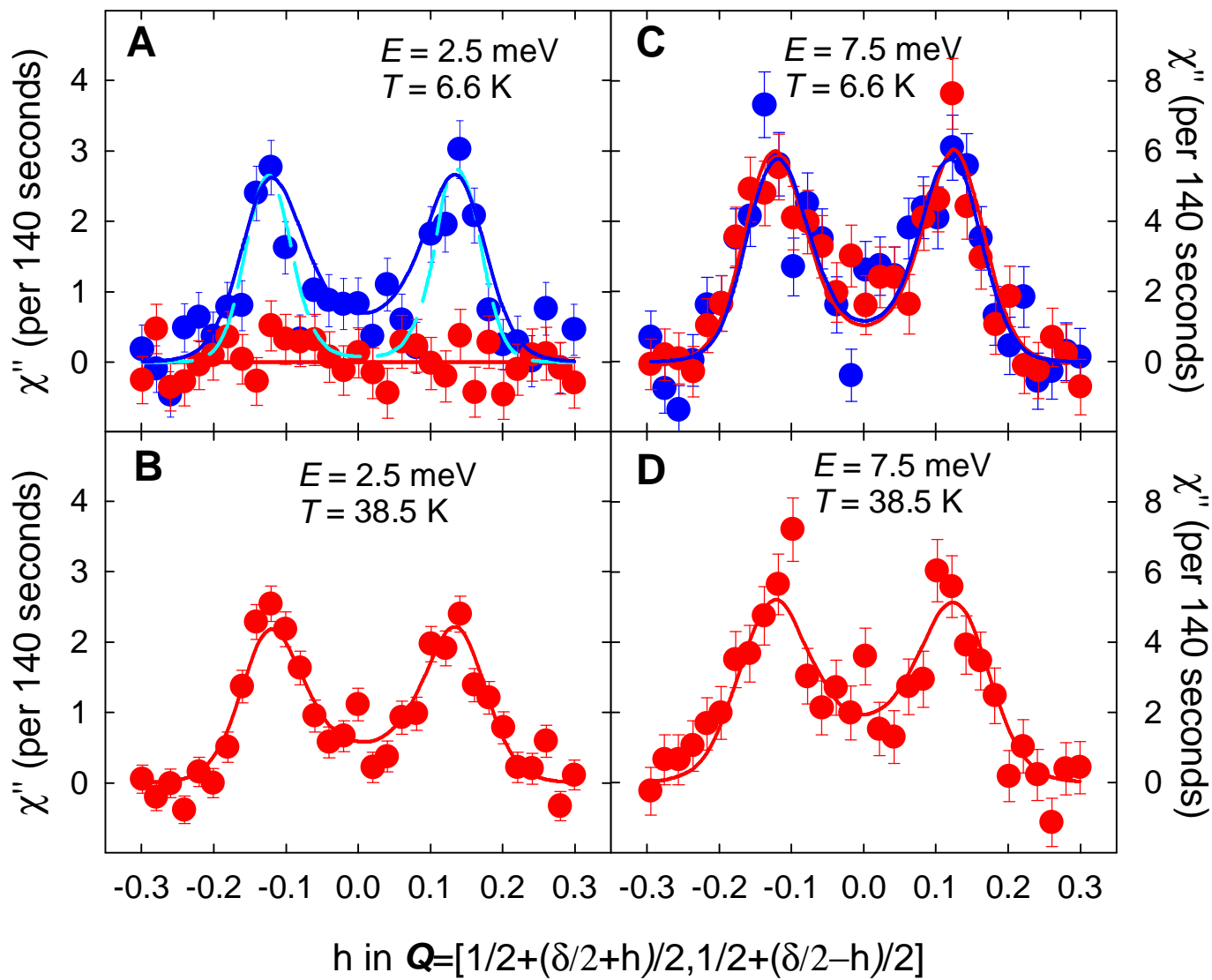
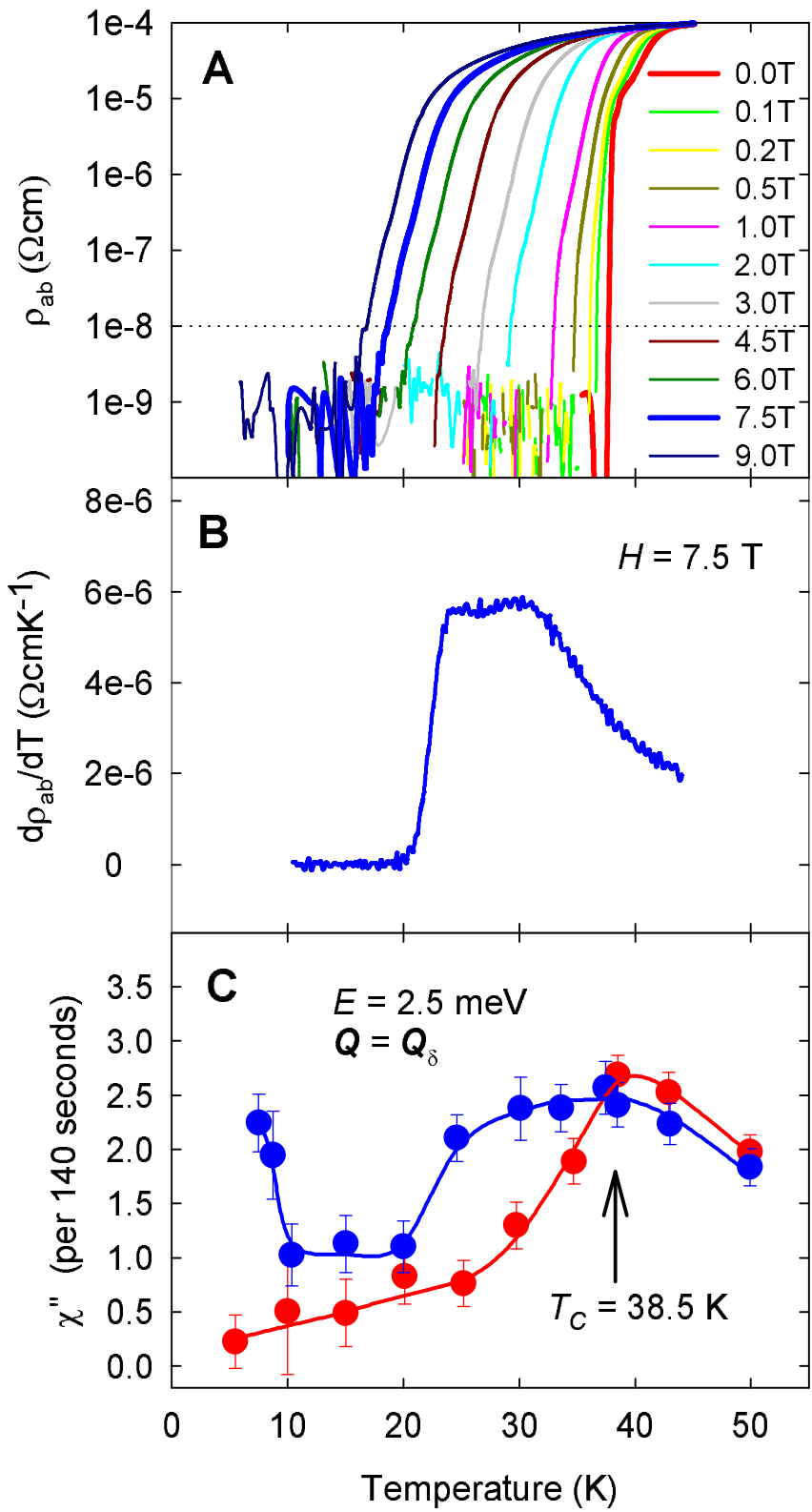
\end{document}